\documentclass[a4paper,12pt]{article} 

\usepackage{cmap}					
\usepackage[T2A]{fontenc}			
\usepackage[utf8]{inputenc}			
\usepackage[english,russian]{babel}	

\usepackage[left=4cm, top=1cm, right=1cm, bottom=20mm, nohead, footskip=12.5mm]{geometry}%

\usepackage{amsmath,amsfonts,amssymb,amsthm,mathtools} 
\usepackage{icomma} 
\usepackage{cancel} 

\addto\captionsrussian{}

\usepackage{euscript}	 
\usepackage{mathrsfs} 

\usepackage{graphicx}  
\graphicspath{{images/}{images2/}}  
\usepackage{wrapfig} 

\usepackage{ esint }

\usepackage{amssymb}
 
\usepackage{amsmath}

\renewcommand\div{{\rm div}\,}
\newcommand\rot{{\rm rot}\,}

\usepackage[usenames]{color}
\usepackage{colortbl}

\usepackage{multicol}
\usepackage{hyperref}
\usepackage{epigraph}
\author{\small \textbf{M.A. Fedotova }$\bf{^{a,b}}$\textbf{, D.A. Klimachkov }$\bf{^a}$\textbf{, A.S. Petrosyan }$\bf{^{a,b}}$\\
\\
\footnotesize $^a$ Space Research Institute of Russian Academy of Science\\
\footnotesize 84/32, Profsoyuznaya str., Moscow, 117997, Russia
\\
\\
\footnotesize $^b$ Moscow Institute of Physics and Technology (State University)\\
\footnotesize 9 Institutskyi per., Dolgoprudny, Moscow Region, 141700, Russia  \normalsize}

\title{Nonlinear theory of magnetohydrodynamic flows of stratified rotating plasma in two-layer shallow water approximation. Rossby waves and their three-wave interactions}
\date{}
\begin{document} 
\maketitle
\selectlanguage{english}
\begin{abstract}
\small This article deals with rotating magnetohydrodynamic flows of a thin stratified layer of astrophysical plasma in a gravitational field with a free-surface in a vertical external magnetic field. Magnetohydrodynamic equations are obtained in the two-layer shallow water approximation in an external magnetic field when plasma is divided into two layers of different densities. In the beta-plane approximation a system of shallow water equations for rotating stratified plasma in an external magnetic field is obtained. For stationary initial conditions in the form of vertical or horizontal magnetic field a linear theory has been developed and solutions have been found in the form of magneto-Rossby waves with modifications to them describing the effects of stratification. A qualitative analysis of the dispersion curves shows the presence of three-wave nonlinear interactions of magneto-Rossby waves for each of the stationary solutions. The appearance of parametric instabilities is shown and their increments are found. \normalsize
\end{abstract}
\tableofcontents
\newpage
\section{Introduction}
The magnetohydrodynamic shallow water theory is important for describing large-scale processes in rotating flows of astrophysical plasma. Shallow water approximation in plasma magnetohydrodynamics is used to describe solar tachocline \cite{1}-\cite{6} atmospheres of exoplanets \cite{7}, atmospheric dynamics of neutron stars \cite{8}-\cite{9}, and accreting matter flows in the neutron stars \cite{9}-\cite{10}. Practically, we deal with the development of ideas of geophysical hydrodynamics for rotating plasma, taking into account the significant differences in the behavior of plasma flows due to the presence of magnetic field.

Flows in plasma astrophysics, as well as flows in geophysics, are usually stratified. This work is devoted to the study of the fundamental role of stratification in astrophysical plasma flows. It should be noted, that the complete system of magnetohydrodynamic equations of a stratified plasma is complicated both for theoretical analysis and for numerical simulations. Useful model for continuous stratification plasma is superimposed n-layers of different densities \cite{12}-\cite{13}. In this article we propose the magnetohydrodynamic equations of stratified plasma in external magnetic field in two-layer shallow water approximation. Equations obtained in \cite{14}-\cite{16} are generalized here to the case of a thin rotating stratified layer of plasma  with a free-surface in an external vertical magnetic field. Two systems of equations are obtained: equations including full Coriolis force and equations in the beta-plane approximation. The simplified magnetohydrodynamic shallow water equations obtained in this work represent the only self-consistent possibility of taking into account the presence of an external vertical magnetic field and stratification. The two-layer magnetohydrodynamic shallow water equations play the same important role in space and astrophysical stratified plasma as the classical shallow water equations in the hydrodynamics of neutral stratified fluid \cite{33}. Accounting for stratification in magnetohydrodynamic models of rotating plasma is important for analysis of R-mode oscillations in rotating stars and in the Sun \cite{17}-\cite{19}, and significantly increases the possibility of interpreting the available observational data for large-scale Rossby waves on the Sun \cite{20}-\cite{23}.

In our work we use the developed two-layer shallow water theory of magnetohydrodynamic flows on beta-plane to study magneto-Rossby waves \cite{15},\cite{24}. Magneto-Rossby waves are large-scale waves arising from the latitudinal inhomogeneities of the Coriolis force for the spherical case. Rossby waves determine the large-scale dynamics of the Sun and stars \cite{25}-\cite{28}, tidally locked magnetoactive atmospheres of exoplanets \cite{7}, and flows in accretion disks and in atmospheres of neutron stars \cite{9}, \cite{29}. In addition, the Rossby waves play a decisive role in the emergence of zonal flows in two-dimensional magnetohydrodynamic turbulence and in Earth's interiors \cite{30}-\cite{32}. Large-scale Rossby waves in a neutral fluid determine the global dynamics of planetary atmospheres and are the subject of numerous studies in geophysical fluid dynamics \cite{33}, \cite{Pokhotelov}, \cite{34}. In the case of astrophysical plasma flows, the theory of Rossby waves is significantly complicated by the presence of a magnetic field; therefore, main results on magneto-Rossby waves are obtained in a linear approximation \cite{3}, \cite{9}, \cite{27}-\cite{28} using the shallow water magnetohydrodynamic theory. Worth of note are important studies on the development of the nonlinear theory of magneto-Rossby waves \cite{15}, \cite{35} as well as magneto-Rossby wave theory for the case of compressible shallow water flows \cite{36}. All the listed phenomena in plasma astrophysics are obtained on the basis of the magnetohydrodynamic shallow water approximation in a plasma without taking into account stratification.

In the present work the dispersion laws of magneto-Rossby waves are obtained both in an external vertical magnetic field and in a horizontal magnetic field taking into account density stratification using the developed two-layer magnetohydrodynamic shallow water theory on beta-plane. It was found that the modifications to magneto-Rossby waves associated with stratification change the waves phase and group velocities. The satisfaction of the phase matching condition for three interacting magneto-Rossby waves is shown and equations of nonlinear interaction are obtained both in the case of the presence of an external magnetic field, and in the case of its absence. Derived coefficients of wave interaction differ from the coefficients in the single layer model equations \cite{15} by the presence of terms related to the difference in densities of plasma layers. The possibility of parametric instabilities is shown and their increments are obtained.
The results obtained in our study for magnetic Rossby waves in the presence of stratification are significant for understanding the dynamics of various astrophysical objects. For example, they allow to detail the wave dynamics of the solar tachocline and the interaction of processes in the tachocline with solar activity and, thus, to advance predictions and analyzing of the solar seasons formation \cite{4}-\cite{5}, \cite{20}, \cite{22}-\cite{23}, \cite{25}.

In section 2, the magnetohydrodynamic equations for rotating stratified plasma are obtained in the two-layer shallow water approximation in an external magnetic field. In section 3, the equations obtained are generalized to the case of spherical flows on beta-plane, and solutions have been found in the form of magneto-Rossby waves with modifications describing effects of stratification. In section 4 it is shown, that the phase matching condition is satisfied for the obtained dispersion relations, the equations of three-wave interaction and the characteristics of parametric instabilities are obtained.
\section{Two-layer magnetohydrodynamic shallow water equations in an external magnetic field.}
Here we derive the magnetohydrodynamic equations describing a stratified plasma in the two-layer shallow water approximation. We take three-dimensional system of magnetohydrodynamic equations for a rotating incompressible plasma in a gravitational field as initial equations:
\begin{equation}\rho\frac{\partial\mathbf{u}}{\partial t}+\rho(\mathbf{u}\nabla)\mathbf{u}=-\nabla p-\frac{[\mathbf{B}\rot\mathbf{B}]}{4\pi}-\rho[\mathbf{f}\mathbf{u}]+\rho \mathbf{g} \label{1.1},\end{equation}
\begin{equation}\frac{\partial\mathbf{B}}{\partial t}=\rot[\mathbf{u}\mathbf{B}]\label{1.2},\end{equation}
\begin{equation}\div\mathbf{u} = 0,\end{equation}
\begin{equation}\div\mathbf{B} = 0,\label{1.4}\end{equation}

In (\ref{1.1}) - (\ref{1.4}) $\vec{u}$ - velocity vector, $\vec{B}$ - magnetic field strength vector, $\rho$ - density, $\mathbf{f}=(0,0,f)$, $f=2\Omega\sin\theta$ - Coriolis parameter, $\Omega$ - angular velocity of rotation, $\theta$ - latitude, $\mathbf{g}=(0,0,-g)$ - gravity acceleration, $p$ - total pressure. The first equation in the system is the momentum equation, the second is the equation for magnetic field, the third is the velocity field divergence-free condition, and the fourth is the magnetic field divergence-free condition. 
Since the free-surface is not strictly horizontal, imposing of vertical magnetic field physically can lead to stresses perpendicular to the surface. Though in present work we use shallow water approximation and we consider horizontal flow scales large comparing with vertical scales. It allows us to neglect all vertical disturbances due to their smallness $\sim h/L$.

We consider the flow of a thin stratified plasma layer with a free-surface in a uniform gravity field, in a rotating frame, in the presence of an external vertical magnetic field $ B_0 $ (Fig. 1).
\begin{figure}[ht]
\center{\includegraphics[scale=0.37]{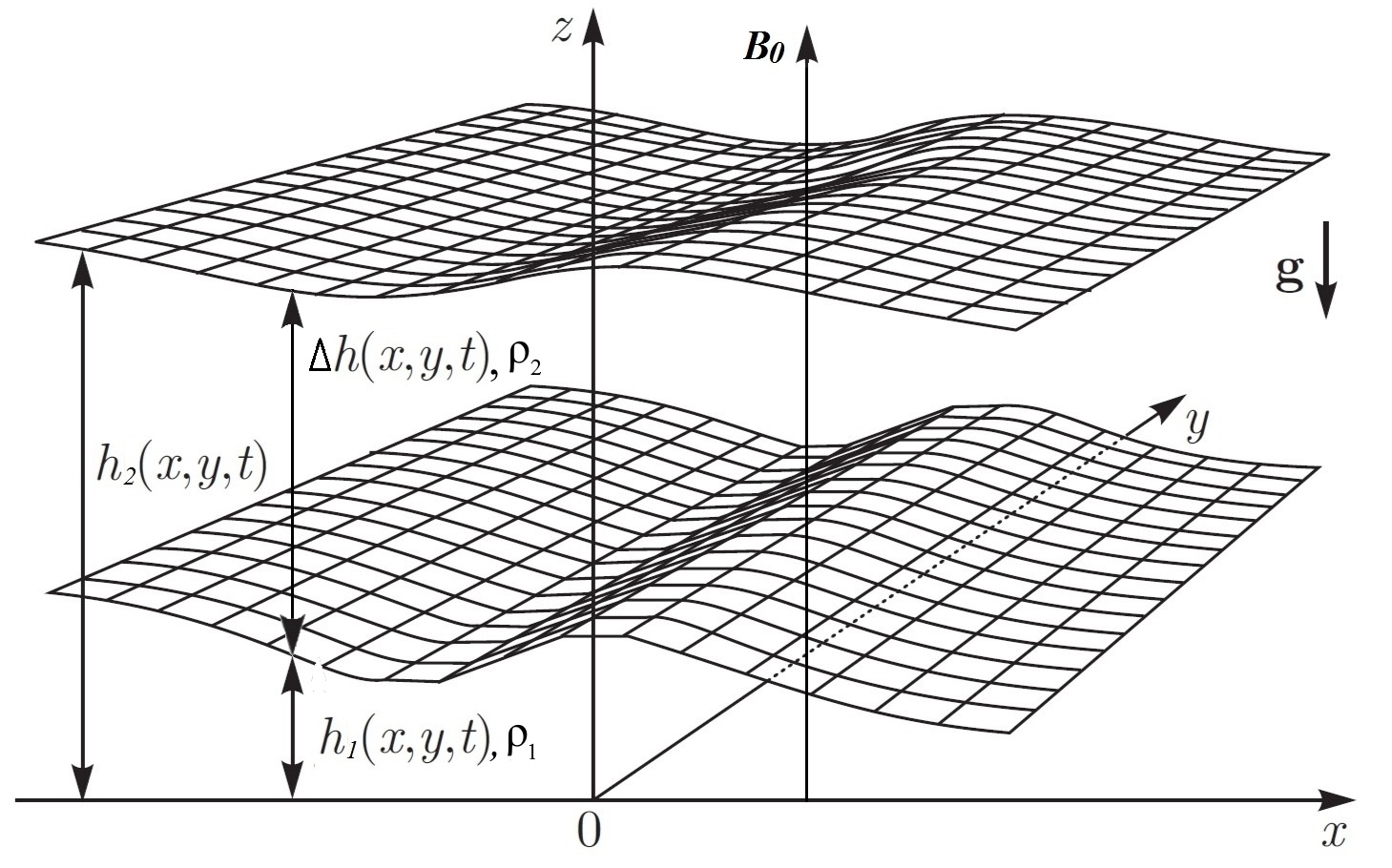}}
\caption{Geometry of the layer.}
\label{ris:image}
\end{figure}

We divide a thin layer of plasma of height $ h_2 $ into two layers: the bottom layer of height $ h_1 $ with a constant density $ \rho_1 $ and the top layer of height $ \Delta h = h_2-h_1 $ with a constant density $ \rho_2 $. Writing down initial system (\ref {1.1}) - (\ref {1.4}) for each of the layers, we integrate it over the height $ h_1 $ for the bottom layer and over the height $ \Delta h $ for the top layer. We consider the heights of each layer to be much less than the characteristic horizontal scales. In this case, neglecting vertical accelerations, the total pressure (the sum of fluid and magnetic pressure) is considered hydrostatic. As a result, magnetohydrodynamic equations are obtained for two layers of plasma with different density in shallow water approximation in an external vertical magnetic field.

Let us rewrite equations (\ref {1.1}) and (\ref {1.2}) for each layer in the matrix form.
\footnotesize
\begin{equation}\label{2}\partial_t\begin{pmatrix} \rho_iu_{1i} \\ \rho_i u_{2i} \\ \rho_iu_{3i} \\ \tilde{B}_{1i} \\ \tilde{B}_{2i} \\ \tilde{B}_{3i} \end{pmatrix}+ \partial_x \begin{pmatrix} \rho_iu_{1i}^2-\tilde{B}^2_{1i}+\tilde{p}_i \\ \rho_i u_{1i}u_{2i} -\tilde{B}_{1i}\tilde{B}_{2i} \\ \rho_1u_{1i}u_{3i}-\tilde{B}_{1i}\tilde{B}_{3i} \\ 0 \\ u_{1i}\tilde{B}_{2i}-u_{2i}\tilde{B}_{1i} \\ u_{1i}\tilde{B}_{3i}-u_{3i}\tilde{B}_{1i}\end{pmatrix}+  \partial_y \begin{pmatrix}  \rho_i u_{1i}u_{2i} -\tilde{B}_{1i}\tilde{B}_{2i} \\ \rho_i u_{2i}^2-\tilde{B}^2_{2i}+\tilde{p}_i \\ \rho_iu_{2i}u_{3i}-\tilde{B}_{2i}\tilde{B}_{3i}  \\ u_{2i}\tilde{B}_{1i}-u_{1i}\tilde{B}_{2i} \\ 0 \\ u_{2i}\tilde{B}_{3i}-u_{3i}\tilde{B}_{2i}\end{pmatrix}
+ \partial_z \begin{pmatrix}  \rho_i u_{1i}u_{3i} -\tilde{B}_{1i}\tilde{B}_{3i} \\ \rho_i u_{2i}u_{3i}-\tilde{B}_{2i}\tilde{B}_{3i} \\\rho_iu^2_{3i}-\tilde{B}^2_{3i} + \tilde{p}_i \\ u_{3i}\tilde{B}_{1i}-u_{1i}\tilde{B}_{3i} \\ u_{31}\tilde{B}_{2i}-u_{2i}\tilde{B}_{3i} \\ 0\end{pmatrix}= \begin{pmatrix} \rho_i fu_{2i} \\ -\rho_i fu_{1i} \\ -\rho_i g \\ 0 \\ 0 \\ 0\end{pmatrix}\end{equation}
\normalsize
In (\ref{2}) $\rho_i$ -- density of the layer, $\tilde{p}_i=p_i+\frac{B_i^2}{8\pi}$ -- magnetohydrostatic pressure in the layer,  $\tilde{B}_{1i}$, $\tilde{B}_{2i}$, $\tilde{B}_{3i}$ -- components of the magnetic field in the bottom layer ($\tilde{\mathbf{B}}^2=\frac{1}{4\pi}\mathbf{B}^2$), $u_{1i}$, $u_{2i}$, $u_{3i}$ -- components of the velocity field in the layer, index $i=1$ corresponds to the bottom layer, and index $i=2$ corresponds to the top layer.\\

Let us write the boundary conditions for each layer of plasma in an external vertical magnetic field. The boundary conditions for the velocity field are following:
\begin{equation} \left.u_3\right|_{z=0}=0 \label{4.1},\end{equation}
\begin{equation}\left.u_3\right|_{z=h_i}=\partial_th_i+\left.u_1\right|_{z=h_i}\partial_xh_i+\left.u_2\right|_{z=h_i}\partial_yh_i\label{4.2}.\end{equation}

At the bottom we use impermeability boundary condition for the velocity (\ref {4.1}). For the boundary between the layers we use the condition of the equality of the vertical components of the velocities in each layer (index i=1 in (\ref {4.2})). The boundary condition on free-surface corresponds to the equality of the vertical component of the velocity of the top layer and the velocity of free-surface (index i=2 in (\ref {4.2})). 

The boundary conditions for the magnetic field are following:
\begin{equation}\left.B_3\right|_{z=0}=B_0\label{5.1},\end{equation}	\begin{equation}\left.B_3\right|_{z=h_i}=\left.B_1\right|_{z=h_i}\partial_xh_i+\left.B_2\right|_{z=h_i}\partial_yh_i+B_0\label{5.2},\end{equation}
here and after $\vec {B} = \tilde {\vec {B}} \frac {1} {\sqrt {\rho}} $.

In the case of absence of the external vertical magnetic field ($B_0=0$ in expressions (\ref{5.1}) - (\ref{5.2})) boundary conditions for the vertical component of the magnetic field $B_3$ on surfaces $z=h_1(x,y)$ and $z=h_2(x,y)$ are defined by parallelism of magnetic induction to this surfaces respectively and are the sum of horizontal components $B_1$ and $B_2$, multiplied by corresponding tangents of angles $\partial h_{i}/\partial x$ and $\partial h_{i}/\partial y$. Therefore, vertical component of the magnetic field is equal to zero on the bottom ($\left.B_3\right|_{z=0}=0$ \ref{5.1}). When we impose external vertical magnetic field ($B_0$) , boundary conditions for vertical component of magnetic field $B_3$ on surfaces $z=0$ (\ref{5.1}), $z=h_i$ (\ref{5.2}) are modified by adding the term $B_0$.\\
\\
Let us write the hydrostatic equation for the total pressure in each layer of plasma:
\begin{equation}\label{4}
\partial_z\left(p_i+\frac{\rho_i}{2}B_i^2\right)=-\rho_i g.
\end{equation}

We use this equation to obtain pressure at the bottom of a thin layer of height $ h_2 $ and at the boundary between plasma layers of different densities, as well as pressure distribution in the layer of height $h_1$ and density $ \rho_1 $ and in the layer of height $ \Delta h $ and density $ \rho_2 $. To do this, let us integrate equation (\ref{4}) over height $h_1$ for the bottom layer and over $\Delta h=h_2-h_1$ for the top layer:
\begin{equation}
\label{davl1}
\int\limits_{a_i}^{h_i}\partial_z\tilde{p}_idz=-\int\limits_{a_i}^{h_i}\rho_igdz,
\end{equation}
here and after index $i=1$ corresponds to the bottom layer, in which $a_i=0$, $h_i=h_1$, and index $i=2$ corresponds to the top layer, in which $a_i=h_1$, $h_i=h_2$.

Assuming the pressure at the free-surface is constant $\left. p\right|_{z=h_2}=p_0$, we find the pressure at the boundary between the layers $\left.\tilde{p}\right|_{h_1}$:
\begin{equation}\label{ph1}
 \left.\tilde{p}\right|_{h_1}=p_0+\rho_2g(h_2-h_1).
\end{equation}

Replacing the upper limit of integration $h_2$ by $z$ in the equation (\ref{davl1}), we find the pressure $\tilde{p}_2(z)$ in the top layer of plasma with density $\rho_2$:
\begin{equation}\label{p2}\tilde{p}_2(z)=p_0+\rho_2g(h_2-z).
\end{equation}

Similarly, we find the pressure at the bottom $\left.\tilde{p}\right|_0$ and the pressure $\tilde{p}_1(z)$ in the bottom layer of plasma with density $\rho_1$ from the equation (\ref{davl1}):
\begin{equation}\label{p0} \left.\tilde{p}\right|_{0}=p_0+\rho_2g(h_2-h_1)+\rho_1gh_1,
\end{equation}
\begin{equation}\label{p1}\tilde{p}_1(z)=p_0+\rho_2g(h_2-h_1)+\rho_1g(h_1-z).
\end{equation}

Leibniz integral rule and the expressions for pressures (\ref{ph1}) - (\ref{p1}) are employed to integrate equations (\ref {2}).

Let us integrate the velocity field divergence-free condition ranging from $0$ to $h_1$ for the bottom layer (index $i=1$) and ranging from $h_1$ to $h_2$ for the top layer (index $i=2$):
\[\frac{\partial}{\partial x}\int\limits_{a_i}^{h_i}u_{1i}dz-\left.u_{1i}\right|_{z=h_i}\frac{\partial h_i}{\partial x}+\left.u_{1i}\right|_{z=a_i}\frac{\partial a_i}{\partial x}+\frac{\partial}{\partial y}\int\limits_{a_i}^{h_i}u_{2i1}dz-\left.u_{2i}\right|_{z=h_i}\frac{\partial h_i}{\partial y}+\left.u_{1i}\right|_{z=a_i}\frac{\partial a_i}{\partial y}+\]
\[+\left.u_{3i}\right|_{z=h_i}-\left.u_{3i}\right|_{z={a_i}}=0.\]

Taking into account boundary conditions (\ref {4.1}) - (\ref {4.2}), we get:
\begin{equation}\label{nepr1}
\frac{\partial (h_i-a_i)}{\partial t}+\frac{\partial }{\partial x}\int\limits_{a_i}^{h_i}u_{1i}dz+\frac{\partial}{\partial y}\int\limits_{a_i}^{h_i}u_{2i}dz=0.
\end{equation}

Similarly, we integrate the divergence-free condition for magnetic field in the bottom and top plasma layers in an external magnetic field and transform them, using the boundary conditions (\ref {5.1}) - (\ref {5.2}) to the following:
\[\frac{\partial }{\partial x}\int\limits_{a_i}^{h_i}B_{1i}dz+\frac{\partial}{\partial y}\int\limits_{a_i}^{h_i}B_{2i}dz=0.\]

Let us integrate the equations for the magnetic field in each layer. Equations for the horizontal components of the magnetic field are following:
\begin{equation}\label{10}
\frac{\partial}{\partial t}\int\limits_{a_i}^{h_i}B_{ji}dz+\frac{\partial}{\partial x}\int\limits_{a_i}^{h_i}(u_{ki}B_{ji}-u_{ji}B_{ki})dz-B_0(\left.u_{ji}\right|_{h_i}-\left.u_{ji}\right|_{a_i})=0,
\end{equation}
indexes $j=1$, $k=2$ correspond to the $x$-component of the magnetic field, and indexes $j=2$, $k=1$ correspond to the $y$-component of the magnetic field; index $i=1$ corresponds to the bottom layer of plasma, where $a_i=0$, and index $i=2$ corresponds to the topl layer of plasma, where $a_i=h_1$.

Equations for the $z$-component of the magnetic field in each layer are following:
\begin{equation}\label{12}
\frac{\partial}{\partial t}\int\limits_{a_i}^{h_i}B_{3i}dz+\frac{\partial}{\partial x}\int\limits_{a_i}^{h_i}u_{1i}B_{3i}dz+\frac{\partial}{\partial y}\int\limits_{a_i}^{h_i}u_{2i}B_{3i}dz-B_0(\left.u_{3i}\right|_{h_i}-\left.u_{3i}\right|_{a_i})=0,
\end{equation}
index $i=1$ corresponds to the bottom layer of plasma, where $a_i=0$, and index $i=2$ corresponds to the topl layer of plasma, where $a_i=h_1$.

Let us do the same to equations for the horizontal velocities in (\ref{2}) in each layer. We integrate momentum equations in each layer, taking into account the boundary conditions (\ref{4.1}) - (\ref{4.2}), (\ref{5.1}) - (\ref{5.2}). Using the expressions for pressures in the bottom layer (\ref{p1}) and at the boundary between the layers of different densities (\ref{ph1}), we get:
\begin{equation*}
\frac{\partial}{\partial t}\int\limits_{a_i}^{h_i}u_{ji}dz+\frac{\partial}{\partial x}\int\limits_{a_i}^{h_i}(u_{ji}^2-B_{ji}^2)dz+\frac{\partial}{\partial y}\int\limits_{a_i}^{h_i}(u_{ji}u_{ki}-B_{ji}B_{ki})dz+\frac{\rho_2}{\rho_i}g(h_i-a_i)\frac{\partial}{\partial x}H_i+
\end{equation*}
\begin{equation}\label{15}+g\frac{\partial}{\partial x}\frac{(h_i-a_i)^2}{2}+B_0B_{ji}=\alpha f\int\limits_{a_i}^{h_i}u_{ki}dz,
\end{equation}
index $i=1$ corresponds to the bottom layer of plasma, where $ a_i=0$, $h_i=h_1$, $H_i=\Delta h$, and index $i=2$ corresponds to the topl layer of plasma, where $a_i=h_1$, $h_i=h_2$, $H_i=h_1$;
indexes $j=1$, $k=2$ and $\alpha=1$ correspond to the $x$-component of the velocity, and indexes $j=2$, $k=1$ and $\alpha=-1$ correspond to the $y$-component of the velocity.

For the final derivation of magnetohydrodynamic two-layer shallow water equations let us introduce the height-averaged velocities $u_{qi}$ and magnetic fields $B_{qi}$ (index $j=1$ corresponds to $q=x$; index $j=2$ corresponds to $q=y$) and represent velocities and magnetic fields in each layer as the sum of the height-averaged values and fluctuations in the following form:
\begin{equation}u_{ji}=u_{qi}+u'_{ji}=\frac{1}{(h_i-a_i)}\int\limits_{a_i}^{h_i}u_{ji}dz+u'_{ji}\label{sr1.1},\end{equation}
\begin{equation}B_{ji}=B_{qi}+B'_{ji}=\frac{1}{(h_i-a_i)}\int\limits_{a_i}^{h_i}B_{ji}dz+B'_{ji}\label{sr1.4},\end{equation}

$u'_{ji}$ -- fluctuations of the velocities in the bottom (index $i=1$) and in the top (index $i=2$) layers. $B'_{ji}$ -- fluctuations of the magnetic fields (index $i=1$) and in the top (index $i=2$) layers.

Let us substitute expressions (\ref{sr1.1})-(\ref{sr1.4}) into equations (\ref{nepr1}), (\ref{10}), (\ref{15}), neglecting terms which include fluctuations \cite{12}, \cite{37}-\cite{39}. As a result we obtain the magnetohydrodynamic equations for stratified plasma in the gravity field in two-layer shallow water approximation in the external magnetic field.
\begin{equation}\label{18bezb}
\left\{
\begin{aligned}
& \partial_t(h_i-a_i)+\partial_x[(h_i-a_i)u_{xi}]+\partial_y[(h_i-a_i)u_{yi}]=0,\\
& \partial_t[(h_i-a_i)u_{xi}]+\partial_x\left[(h_i-a_i)\left(u_{xi}^2-B_{xi}^2+\frac{g(h_i-a_i)}{2}\right)\right]+\frac{\rho_2}{\rho_i}g(h_i-a_i)\partial_xH_i+\\
& + \partial_y[(h_i-a_i)(u_{xi}u_{yi}-B_{xi}B_{yi})]+B_0B_{xi}=(h_i-a_i)fv_{yi},\\
& \partial_t[(h_i-a_i)u_{yi}]+\partial_x[(h_i-a_i)(u_{xi}u_{yi}-B_{xi}B_{yi})]+\partial_y\left[(h_i-a_i)\left(u_{yi}^2-B_{yi}^2+\frac{g(h_i-a_i)}{2}\right)\right]+\\
&+\frac{\rho_2}{\rho_i}g(h_i-a_i)\partial_yH_i+B_0B_{yi}=-(h_i-a_i)fu_{xi},\\
& \partial_t[(h_i-a_i)B_{xi}]+\partial_y[(h_i-a_i)(B_{xi}u_{yi}-B_{yi}u_{xi})]-B_0u_{xi}=0 ,\\
& \partial_t[(h_i-a_i)B_{yi}]+\partial_x[(h_i-a_i)(B_{yi}u_{xi}-B_{xi}u_{yi})]-B_0u_{yi}=0 ,\\
& \partial_tB_{zi}+B_0(\partial_xu_{xi}+\partial_yu_{yi})=0,\\
& \partial_xB_{xi}+\partial_yB_{yi}=0,\end{aligned}
\right.
\end{equation}

where index $i=1$ corresponds to the bottom layer, where $a_i=0$, $h_i=h_1$, $H_i=\Delta h$, and index $i=2$ corresponds to the top layer, where $a_i=h_1$, $h_i=h_2$, $H_i=h_1$. The first equation -- the equation which describes the variation of the hight of each layer of plasma. The second and the third equations -- equations for the hight-averaged horizontal velocities. The fourth and the fifth equations -- equations for the hight-averaged horizontal magnetic fields.\\

It should be noted that the presence of a vertical magnetic field leads to significant changes in the horizontal dynamics of the magnetic field in the shallow water approximation \cite {14}. The equations for hight ($h_i-a_i$), horizontal velocities ($u_{xi}$, $u_{yi}$) and horizontal magnetic fields ($B_{xi}$, $B_{yi}$) are closed set of equations, which are used for the following research. Last two equations in (\ref{18bezb}) in provide the magnetic field divergence-free condition.In addition these equations describe the fundamental three-dimensionality and axisymmetry of magnetic fields in the shallow water approximation. When imposed vertical magnetic field $ B_0 = 0 $, equations (\ref {18bezb}) are transformed into magnetohydrodynamic equations of a stratified plasma in the two-layer shallow water approximation, obtained in  \cite{12}, \cite{13}. When heights and densities of the layers are equal, equations (\ref {18bezb}) are transformed into magnetohydrodynamic equations in the single-layer shallow water approximation in the external magnetic field  \cite {14}, and at $ B_0 = 0 $ are reduced to well-known shallow water magnetohydrodynamic equations without an external magnetic field \cite{1}, \cite{12}, \cite{40}, \cite{41}.

\section{The beta-plane approximation. Rossby waves}
Let us research below spherical flows of a thin layer of incompressible rotating plasma in the two-layer shallow water approximation within the obtained equations (\ref{18bezb}). The effects of sphericity we are taking into account in in the beta-plane approximation by analogy with the equations of neutral fluid \cite {14}. It is assumed that the Coriolis parameter $f$ varies only slightly with small changes in latitude. Let us express $f$ in the following form:
\begin{equation}\label{beta}f=2\Omega\sin{\theta}\approx 2\Omega\sin{\theta_0}+2\Omega(\theta-\theta_0)\cos{\theta_0}\approx f_0+\beta y,\end{equation}
$\Omega$ - angular velocity of rotation, which is equal for both layers, $f_0=2\Omega\sin{\theta_0}$, $\beta=\partial f/\partial y$, $y$ coordinate is measured along latitude in the north direction and related to $ \ theta $ as follows $y=r(\theta-\theta_0)$, $r$ -- radius of the sphere.

In system (\ref {18bezb}) we take derivatives of equations for horizontal components $ u_{xi} $  of velocity with respect to $ y $ taking into account the dependence of the Coriolis parameter on latitude (\ref {beta}). Considering $ \beta y \ll f_0 $, we obtain the magnetohydrodynamic equations of a stratified plasma in a gravity field in the two-layer shallow water approximation in the external magnetic field on the beta-plane:
\begin{equation}\label{18}
\left\{
\begin{aligned}
& \partial_t(h_i-a_i)+\partial_x[(h_i-a_i)u_{xi}]+\partial_y[(h_i-a_i)u_{yi}]=0,\\
& \partial_y\partial_t[(h_i-a_i)u_{xi}]+\partial_y\partial_x\left[(h_i-a_i)\left(u_{xi}^2-B_{xi}^2+\frac{g(h_i-a_i)}{2}\right)\right]+\frac{\rho_2}{\rho_i}g\partial_y(h_i-a_i)\partial_xH_i+\\
&+\partial^2_y[(h_i-a_i)(u_{xi}u_{yi}-B_{xi}B_{yi})]+B_0\partial_yB_{xi}=f_0\partial_y[(h_i-a_i)v_{yi}]+\beta (h_i-a_i)v_{yi},\\
& \partial_t[(h_i-a_i)u_{yi}]+\partial_x[(h_i-a_i)(u_{xi}u_{yi}-B_{xi}B_{yi})]+\partial_y\left[(h_i-a_i)\left(u_{yi}^2-B_{yi}^2+\frac{g(h_i-a_i)}{2}\right)\right]+\\
&+\frac{\rho_2}{\rho_i}g(h_i-a_i)\partial_yH_i+B_0B_{yi}=-(h_i-a_i)f_0u_{xi}, \\
& \partial_t[(h_i-a_i)B_{xi}]+\partial_y[(h_i-a_i)(B_{xi}u_{yi}-B_{yi}u_{xi})]-B_0u_{xi}=0 ,\\
& \partial_t[(h_i-a_i)B_{yi}]+\partial_x[(h_i-a_i)(B_{yi}u_{xi}-B_{xi}u_{yi})]-B_0u_{yi}=0 ,\end{aligned}
\right.
\end{equation}
where index $i=1$ corresponds to the bottom layer, where $a_i=0$, $h_i=h_1$, $H_i=\Delta h$, and index $i=2$ corresponds to the top layer, where $a_i=h_1$, $h_i=h_2$, $H_i=h_1$. The first equation -- the equation which describes the variation of the hight of each layer of plasma. The second and the third equations -- equations for the hight-averaged horizontal velocities in beta-plane approximation for the Coriolis force. The fourth and the fifth equations -- equations for the hight-averaged horizontal magnetic fields.
Waves driven by the latitudinal dependence of the Coriolis force are usually called magneto-Rossby waves \cite{15} by analogy to Rossby waves in the neutral fluid dynamics. 

We use the obtained equations (\ref {18}) to study magneto-Rossby waves in stratified plasma in an external vertical magnetic field in two-layer shallow water approximation on beta-plane. In the absense of external vertical magnetic field, the equations obtained are transformed to the two-layer magnetohydrodynamic shallow-water equations on the beta plane and are used below to study magneto-Rossby waves in toroidal and poloidal magnetic fields.

\subsection{Linear magneto-Rossby waves in an external vertical magnetic field}
We consider the flow of a thin stratified layer of plasma in the shallow water approximation on the beta-plane in an external vertical magnetic field.

Let us linearize the equations (\ref {18}) with respect to the stationary solution:
\[h_{1,2}=h_{01,02}=\mathsf{const}\textsf{;    }u_{x1}=u_{y1}=u_{x2}=u_{y2}=B_{x1}=B_{y1}=B_{x2}=B_{y2}=0\textsf{; }B_0=\mathsf{const}\]

From the condition that the determinant of the matrix of the linearized system is zero, we obtain the following dispersion relation for waves in a rotating stratified plasma in an external field in two-layer shallow water approximation on beta-plane:
\begin{equation}\label{23}
(\omega^4-b_1\omega^2-c_1\omega+d_1)(\omega^4-b_2\omega^2-c_2\omega+d_2)=\frac{\rho_2}{\rho_1}g^2k^4h_{01}\Delta h_0(\omega^2+q'\omega+q_1)(\omega^2+q'\omega+q_2),
\end{equation}
\[b_j=\frac{2B_0^2}{(h_{0j}-a_{0j})^2}+f_0^2+gk^2(h_{0j}-a_{0j})\textsf{;	}c_j=\beta gk_x(h_{0j}-a_{0j});\]
\[d_j=\frac{B_0^4}{(h_{0j}-a_{0j})^4}+\frac{B_0^2gk^2}{(h_{0j}-a_{0j})}\textsf{;	}q'=\frac{\beta k_x}{k^2}\textsf{;	}q_j=\frac{B_0^2}{(h_{0j}-a_{0j})^2},\]\\
The right-hand side of the dispersion relation (\ref {23}) describes the effects of stratification in a two-layer model and the left-hand side is the product of two expressions. The first one corresponds to the bottom layer and the second one to the top. Strong theoretical analysis of the obtained dispersion equation (\ref {23}) is not possible. We confine ourselves to a qualitative consideration. In the first approximation, we select the Rossby waves in the absence of the stratification \cite{15}. In the case of small differences in densities of plasma layers, we represent the solution of the dispersion equation (\ref {23}) as the sum of the magneto-Rossby wave without stratification and a small modification related to plasma stratification.

Let us find a solution for the magneto-Rossby wave in an external vertical magnetic field in the absence of stratification in the system. Equation (\ref {23}) with $ \rho_1 = \rho_2 $ becomes \\
\begin{eqnarray}\label{24}
[\omega^4-\omega^2(\frac{B_0^2}{h_{01}^2}+\frac{B_0^2}{\Delta h_0^2}+f_0^2)+\frac{B_0^4}{h_{01}^2\Delta h_0^2}][\omega^4-\omega^2(\frac{B_0^2}{h_{01}^2}+\frac{B_0^2}{\Delta h_0^2}+f_0^2+gk^2H)-\omega gH\beta k_x +\nonumber\\ +\frac{B_0^2}{h_{01}\Delta h_0}(\frac{B_0^2}{h_{01}\Delta h_0}+gk^2\frac{h_{01}^3+\Delta h_0^3}{h_{01}\Delta h_0})]=0,
\end{eqnarray}
\\
and we get the following expression for the magneto-Rossby wave in the absence of stratification: \\
\begin{equation}\label{25}
\omega_{MR_1}\approx \frac{\frac{B_0^2}{h_{01}\Delta h_0}(\frac{B_0^2}{h_{01}\Delta h_0}+\frac{gk^2(h_{01}^3+\Delta h_0^3)}{h_{01}\Delta h_0})}{\beta k_xgH}
\end{equation}

It should be noted, that the expression for $ \omega_ {MR_1} $ includes the heights of both layers explicitly. For equal heights of the layers $ h_{01} = \Delta h_0 = H / 2 $, the expression (\ref {25}) describes the Rossby wave in a single-layer approximation \cite {15}:
\[\omega_{MR_1}'\approx\frac{4\frac{B_0^2}{H^2}\left(4\frac{B_0^2}{H^2}+gk^2H\right)}{\beta k_xgH}\]
\\

Let us find out the modification to the frequency associated with the stratification ($ \rho_1 \neq \rho_2 $). We rewrite the equation (\ref {23}) in the following form:\begin{eqnarray}\label{26}[\omega^4-\omega^2(\frac{B_0^2}{h_{01}^2}+\frac{B_0^2}{\Delta h_0^2}+f_0^2)+\frac{B_0^4}{h_{01}^2\Delta h_0^2}][\omega^4-\omega^2(\frac{B_0^2}{h_{01}^2}+\frac{B_0^2}{\Delta h_0^2}+f_0^2+gk^2h_{02})-\omega gh_{02}\beta k_x +\nonumber\\
+\frac{B_0^2}{h_{01}\Delta h_0}(\frac{B_0^2}{h_{01}\Delta h_0}+gk^2\frac{h_{01}^3+\Delta h_0^3}{h_{01}\Delta h_0})]=\left(\frac{\rho_2}{\rho_1}-1\right)g^2k^4h_{01}\Delta h_0\cdot\nonumber\\
\cdot(\omega^2+\frac{\beta k_x}{k^2}\omega+\frac{B_0^2}{h_{01}^2})(\omega^2+\frac{\beta k_x}{k^2}\omega+\frac{B_0^2}{\Delta h_0^2}).\end{eqnarray}

We suppose the modification $ \delta_1 = \omega- \omega_ {MR_1} $ is small compared to the frequency $ \omega_ {MR_1} $. Denote the expression on the right-hand side as $ \varphi_1 \left(\frac {\rho_2}{\rho_1}, \omega_{MR_1}\right)$.

When the expression in the first bracket of (\ref {26}) is equal to zero, we get the following expressions for the squared frequency:
\[\omega_{1,2}^2=\frac{1}{2}\left(f_0^2+\frac{B_0^2}{h_{01}^2}+\frac{B_0^2}{\Delta h_0^2}\pm\sqrt{\left(\frac{B_0^2}{h_{01}^2}-\frac{B_0^2}{\Delta h_0^2}\right)^2+f_0^2\left(f_0^2+\frac{2B_0^2}{h_{01}^2}+\frac{2B_0^2}{\Delta h_0^2}\right)}\right)\]
If the expression in the first bracket of (\ref {26}) is not zero, we find the modification to the magneto-Rossby wave in an external vertical field related to the presence of stratification:
\begin{equation}\label{27}
\delta_1 =-\frac{\varphi_1\left(\frac{\rho_2}{\rho_1},\omega_{MR_1}\right)}{(\omega_{MR_1}^2-\omega^2_1)(\omega_{MR_1}^2-\omega_2^2)gh_{02}\beta k_x}\end{equation}

We write the phase $ v_ {ph_ {x_1}} $ and group $ v_ {gr_ {x_1}} $ velocities in the $ k_x $ direction for the obtained Rossby wave in the model of two layers of different density (\ref{25}),(\ref{27}):
\begin{eqnarray}\label{vph1}
v_{ph_{x_1}}=\frac{\omega_{MR_1}+\delta_1}{k_x}=\frac{B_0^2(B_0^2+gk^2(h_{01}^3+\Delta h_0^3))}{h_{01}^2\Delta h_0^2h_{02}\beta g k_x^2}+\frac{-\varphi_1\left(\frac{\rho_2}{\rho_1},\omega_{MR_1}\right)}{(\omega_{MR_1}^2-\omega^2_1)(\omega_{MR_1}^2-\omega_2^2)gh_{02}\beta k_x^2},\end{eqnarray}
\begin{eqnarray}\label{vgr1}
v_{gr_{x_1}}=\frac{\partial(\omega_{MR_1}+\delta_1)}{\partial k_x}=-\frac{B_0^2(B_0^2+g(h_{01}^3+\Delta h_0^3)(k_y^2-k_x^2)}{h_{01}^2\Delta h_0^2h_{02}\beta g k_x^2}+\nonumber\\
+\frac{\partial}{\partial k_x}\left(\frac{-\varphi_1\left(\frac{\rho_2}{\rho_1},\omega_{MR_1}\right)}{(\omega_{MR_1}^2-\omega^2_1)(\omega_{MR_1}^2-\omega_2^2)gh_{02}\beta k_x}\right).\end{eqnarray}

Expressions (\ref {vph1}), (\ref {vgr1}) show that the presence of a stratification ($ \rho_2 \neq \rho_1 $) in the system increases the phase velocity (\ref {vph1}) of the magneto-Rossby wave along the $ k_x $ in a vertical magnetic field and reduces its group velocity (\ref {vgr1}) in this direction. 

It should be noted that the dispersion equation (\ref {24}) in the absence of an external magnetic field reduces to the dispersion equation for the neutral fluid layer of height $ h_ {02} $ in the shallow water approximation \cite {15}:
\begin{equation}\label{28}(\omega^2-f_0^2)(\omega^3-\omega(f_0^2+gk^2h_{02})-gk_x\beta h_{02})=0,\end{equation}

with the solution in form of hydrodynamic Rossby wave:
\begin{equation}\label{29}\omega_R=-\frac{gk_x\beta h_{02}}{f_0^2+gk^2h_{02}}\end{equation}

Similarly, we find out the modification related to stratification for the hydrodynamic Rossby wave. Dispersion equation with small difference in density has the form:
\begin{equation}\label{30}(\omega^2-f_0^2)(\omega^3-\omega(f_0^2+gk^2h_{02})-gk_x\beta h_{02})=\left(\frac{\rho_2}{\rho_1}-1\right)g^2k^4h_{01}\Delta h_0\left(\omega+\frac{2\beta k_x}{k^2}\right)\end{equation}
Denoting the expression on the right-hand side as $\xi\left(\frac{\rho_2}{\rho_1}, \omega_R\right)$ and desired modification as $\delta_N=\omega-\omega_R$, we obtain the following expression, taking into account that $\omega_R^2\neq f_0^2$:
\begin{equation}\label{31}\delta_N=\frac{\xi(\frac{\rho_2}{\rho_1},\omega_R)}{(f_0^2+gk^2h_{02})(\omega_R^2-f_0^2)}\end{equation}\\

Thus, it has been shown that in the linear approximation, the system of equations of two-layer shallow water in an external vertical magnetic field (\ref {18}) has a solution in the form of a magneto-Rossby wave. The dependence of the dispersion equation on the ratio of the plasma layer densities is obtained. It is found that the frequency modification associated with stratification reduces the group velocity of the magnetic Rossby waves in an external vertical magnetic field $ v_{gr_{x1}} $ and increases the phase velocity $ v_{ph_{x1}} $. Let us note that the parameter $ \beta $, which describes the effects of sphericity, is present both in the expression for the frequency of the magneto-Rossby wave in an external vertical magnetic field without taking into account the stratification $ \omega_{MR_1} $ (\ref {25}), and in for the $ \delta_1 $ (\ref {27}) related to stratification. However, for the hydrodynamic Rossby wave, the parameter $ \beta $ is absent in the $ \delta_N $ correction (\ref {31}) associated with the stratification.
\subsection{Linear magneto-Rossby waves in horizontal magnetic field}
We now consider the flow of a thin stratified layer of plasma in two-layer shallow water approximation on beta-plane in the absence of an external vertical magnetic field. In this case equations (\ref {18}) have a stationary solution in the form of a horizontal magnetic field:
\[\vec{u}=0\textsf{;	}h_1=h_{01}=\mathsf{const}\textsf{;	}h_2=h_{02}=\mathsf{const}\textsf{;	}\vec{B}=\vec{B}_0=\mathsf{const}\]

From the condition that the determinant of the matrix of the linearized system is zero, we obtain the following dispersion relation for waves in a rotating stratified plasma in horizontal magnetic field in two-layer shallow water approximation on beta-plane:
\begin{equation}\label{33}
(\omega^4-b_1\omega^2-c_1\omega +d_1)(\omega^4-b_2\omega^2-c_2\omega +d_2)=\frac{\rho_2}{\rho_1}g^2k^4h_{01}\Delta h(\omega^2+q\omega-p_1)(\omega^2+q\omega-p_2),
\end{equation}
\[b_j=f_0^2+2(k;B)_j^2+gk^2(h_{0j}-a_{0j})\textsf{	;	}c_j=g(h_{0j}-a_{0j})\beta k_x;\]
\[d_j=(k;B)_j^2((k;B)_j^2+gk^2(h_{0j}-a_{0j});\]
\[q=\frac{\beta k_x}{k^2}\textsf{	;	}p_j=(k;B)_j^2\textsf{	;	}(k;B)_j^2=(k_xB_{x0j}+k_yB_{y0j})^2.\]\\
The right-hand side of the dispersion relation (\ref {33}) describes the effects of stratification in a two-layer model, the left-hand side is the product of two expressions. The first corresponds to the bottom layer, and the second to the top. Strong theoretical analysis of the obtained dispersion equation (\ref {33}) is not possible. We confine ourselves to a qualitative consideration. In the first approximation, we select Rossby waves in the absence of the stratification \cite{15}. In the case of small differences in densities of plasma layers, we represent a solution of the dispersion equation (\ref {33}) as sum of the magneto-Rossby wave in the absence of stratification and a small modification related to plasma stratification.

Let us find a particular solution to the dispersion equation (\ref {33}) for the case of equal magnetic fields in layers $ (k;B)_1 = (k;B)_2 \equiv (k; B) $ in the form of magneto-Rossby wave in the absence of stratification in the system ($ \rho_1 = \rho_2 $). Thus equation (\ref {33}) takes form:
\begin{eqnarray}\label{34}
(\omega^4-\omega^2(f_0^2+2(k;B)^2)+(k;B)^4)(\omega^4-\omega^2(f_0^2+2(k;B)^2+gh_{02}k^2)-\omega gh_{02}\beta k_x+\nonumber\\
+(k;B)^2((k;B)^2+gh_{02}k^2))=0.\end{eqnarray}
The first bracket in the equation (\ref {34}) has a form of the dispersion equation for a thin layer of plasma of height $ h_{02} $ in the one-layer shallow water approximation on beta-plane:
\[\omega^4-\omega^2(f_0^2+2(k;B)^2+gh_{02}k^2)-\omega gh_{02}\beta k_x+ (k;B)^2((k;B)^2+gh_{02}k^2)=0,\]
with solution in form of magneto-Rossby wave: \\
\begin{equation}\label{35}
\omega_{MR_2}\approx \frac{(k;B)^2((k;B)^2+gk^2h_{02})}{\beta k_xgh_{02}}\end{equation}

In particular case of toroidal magnetic field, the dispersion relation (\ref {35}) takes the form:
\[\omega_{MR_2x}\approx \frac{k_xB_x^2(k_x^2B_x^2+gk^2h_{02})}{\beta gh_{02}}\]
It should be noted that $ \omega_ {MR_2} $ does not differ from the Rossby wave in the layer of plasma with height $h_{02}$ \cite{15}.

Let us find the modification to the frequency associated with stratification ($ \rho_1 \neq \rho_2 $). Equation (\ref {33}) takes form:
\begin{eqnarray}\label{36}
(\omega^4-\omega^2(f_0^2+2(k;B)^2)+(k;B)^4)(\omega^4-\omega^2(f_0^2+2(k;B)^2+gh_{02}k^2)-\nonumber\\
-\omega gh_{02}\beta k_x+(k;B)^2((k;B)^2+gh_{02}k^2))=\left(1-\frac{\rho_2}{\rho_1}\right)g^2k^2h_{01}\Delta h_0\cdot\nonumber\\
\cdot\left[k^2\omega^4-2\beta k_x\omega^3+2(k;B)^2k^2\omega^2+2(k;B)^2\beta k_x\omega-(k;B)^4.\right]
\end{eqnarray}
We suppose the modification $ \delta_2 = \omega-\omega_{MR_2} $ is small compared to the frequency $ \omega_ {MR_2} $. The expression on the right-hand side of (\ref{33}) is denoted as $ \varphi_2 \left(\frac {\rho_2} {\rho_1}, \omega_ {MR_2} \right) $.
When the first bracket of the dispersion relation (\ref {36}) is equal to zero, we get following expressions for squared frequency:
\[\omega_{3,4}^2=\frac{f_0^2}{2}+(k;B)^2\pm f_0\sqrt{\frac{f_0^2}{4}+(k;B)^2}\]
If the first bracket of (\ref {36}) is not equal to zero, we obtain the modification to the frequency in the following form:
\begin{equation}\label{37}\delta_2 =-\frac{\varphi_2\left(\frac{\rho_2}{\rho_1},\omega_{MR_2}\right)}{(\omega_{MR_2}^2-\omega_3^2)(\omega_{MR_2}^2-\omega_4^2)gh_{02}\beta k_x}\end{equation}

We write phase $ v_ {ph_ {x_2}} $ and group $ v_ {gr_ {x_2}} $ velocities in the $ k_x $ direction for the obtained Rossby wave in the model of two layers of different density (\ref{35}), (\ref{37}):
\begin{eqnarray}\label{vph2}
v_{ph_{x_2}}=\frac{\omega_{MR_2}+\delta_2}{k_x}=\frac{(k;B)^2((k;B)^2+gk^2h_{02})}{h_{02}\beta g k_x^2}+\frac{-\varphi_2\left(\frac{\rho_2}{\rho_1},\omega_{MR_2}\right)}{(\omega_{MR_2}^2-\omega^2_3)(\omega_{MR_2}^2-\omega_4^2)gh_{02}\beta k_x^2},\end{eqnarray}
\begin{eqnarray}\label{vgr2}
v_{gr_{x_2}}=\frac{\partial(\omega_{MR_2}+\delta_2)}{\partial k_x}=\frac{1}{h_{02}\beta g k_x^2}\left[(k;B)((k;B)^2(4B_{x0}k_x-(k;B))+\right.\nonumber\\
+gh_{02}(2B_{x0}k_x(k_x^2+k_y^2)+\left.+2(k;B)k_x^2-(k;B)(k_x^2+k_y^2))\right]+\nonumber\\
+\frac{\partial}{\partial k_x}\left(\frac{-\varphi_2\left(\frac{\rho_2}{\rho_1},\omega_{MR_2}\right)}{(\omega_{MR_2}^2-\omega^2_3)(\omega_{MR_2}^2-\omega_4^2)gh_{02}\beta k_x}\right).\end{eqnarray}

Expressions (\ref {vph2}), (\ref {vgr2}) show that the presence of stratification ($ \rho_2 \neq \rho_1 $) in the system reduces the group velocity (\ref {vgr2}) of the magneto-Rossby wave along the $ k_x $. Phase velocity of magneto-Rossby wave in the $ k_x $ (\ref {vph2}) direction for very small $ k_x $ ($ k_x <1 $) increases with difference in densities. However, for $ k_x> 1 $ the presence of stratification in the system leads to a noticeable decrease in phase velocity of the wave (\ref {vph2}) along $ k_x $. 

If densities in the dispersion relation (\ref {33}) are equal to each other $ \rho_1 = \rho_2 $ and the external magnetic field is equal to zero $ B_0 = 0$, equation (\ref{33}) reduces to the dispersion equation for neutral fluid layer of height $ h_{02} $ ( \ref {28}) with a solution in the form of a hydrodynamic Rossby wave (\ref {29}) and modification to it in the form (\ref {31}).

Thus, it has been shown that in the absence of an external vertical magnetic field, the system (\ref {18}) in the linear approximation has a solution in the form of a magnetic Rossby wave in a horizontal magnetic field modified by the plasma layer density ratio. The modifications of the frequency related to stratification change the group $ v_{gr_ {x2}} $ and phase $ v_{ph_ {x2}} $ velocities of the Rossby wave. Let us note that the dispersion relations obtained in the section for the frequency of the magneto-Rossby wave in the horizontal magnetic field $ \omega_{MR_1} $ and the modification associated with the $ \delta_1 $ stratification differ significantly from similar expressions obtained for the plasma in an external magnetic field $ \omega_ {MR_2} $, $ \delta_2 $. The parameter $ \beta $, which describes the effects of sphericity, is also present in the expression for the frequency of the magnetic Rossby wave in the horizontal magnetic field $ \omega_ {MR_1} $ (\ref {35}) and in the expression for the modification to it $ \delta_1 $, related to stratification (\ref {37}), as was noted in the case of an external vertical magnetic field.
\section{Three-waves interactions of magneto-Rossby waves. Parametric instabilities}
In this section we study weakly nonlinear interactions of magneto-Rossby waves in two-layer shallow water model. In order to estimate the possibility of wave interactions, let us analyze the dispersion relations obtained in Section 3. The phase matching conditions for three interacting waves with wave vectors $\vec{k}_1$, $\vec{k}_2$ and $\vec{k}_3$ and frequencies $\omega(\vec{k}_1)$, $\omega(\vec{k}_2)$ and $\omega(\vec{k}_3)$ respectively is \cite{15}:
\begin{equation}\label{38}
\omega(\vec{k}_1)+\omega(\vec{k}_2)=\omega(\vec{k}_1+\vec{k}_2)\textsf{;	}\vec{k}_1+\vec{k}_2=\vec{k}_3\end{equation}
To determine whether three magneto-Rossby waves exist both in an external magnetic field (\ref {25}), (\ref {27}) and in a horizontal field (\ref {35}), (\ref {37}), according to the phase matching condition (\ref {38}), we imagine two dispersion curves on the plane ($\omega,k_x$) (Fig. 2) shifted from each other for each case (the presence of an external magnetic field is on the left, the absence of an external magnetic field is on the right). The first term $ \omega (\vec {k} _1) $ in the phase matching condition (\ref {38}) specifies the point ($ k_1, \omega (k_1) $) on the dispersion curve (1) of the solutions of one solution. The second term $ \omega (\vec {k}_2) $ in (\ref{38}) specifies the point ($ k_2, \omega (k_2) $) on the dispersion curve (2) of the solutions of another solution. The phase matching condition (\ref{38}) is satisfied when the second dispersion curve (2) shifted by ($ k_1, \omega (k_1) $) crosses the first one (1) in the point ($ k_3, \omega (k_3) $).

As can be seen from the figures ((Fig. 2), (Fig. 3)), both in the case of an external magnetic field and in the case of a horizontal field, the phase matching condition is satisfied \cite{42}. 

\begin{figure}[h]
\begin{center}
\begin{minipage}[h]{0.45\linewidth}
\[\includegraphics[scale=0.25]{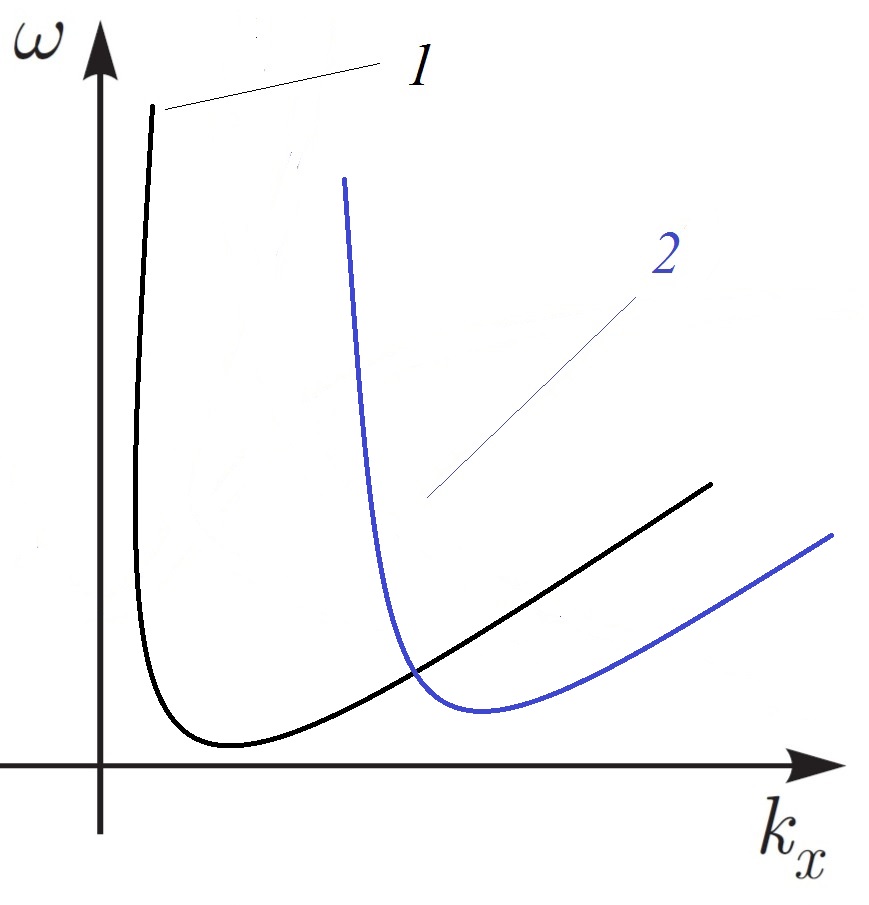}\]
\caption{The phase matching condition for magneto-Rossby waves in an external magnetic field ($B_0\neq0$), 1: $\omega=\omega(k)$,2: $\omega=\omega(k-k_{x_1})-\omega(k_{x_1})$} 
\label{ris:experimoriginal} 
\end{minipage}
\hfill 
\begin{minipage}[h]{0.45\linewidth}
\[\includegraphics[scale=0.25]{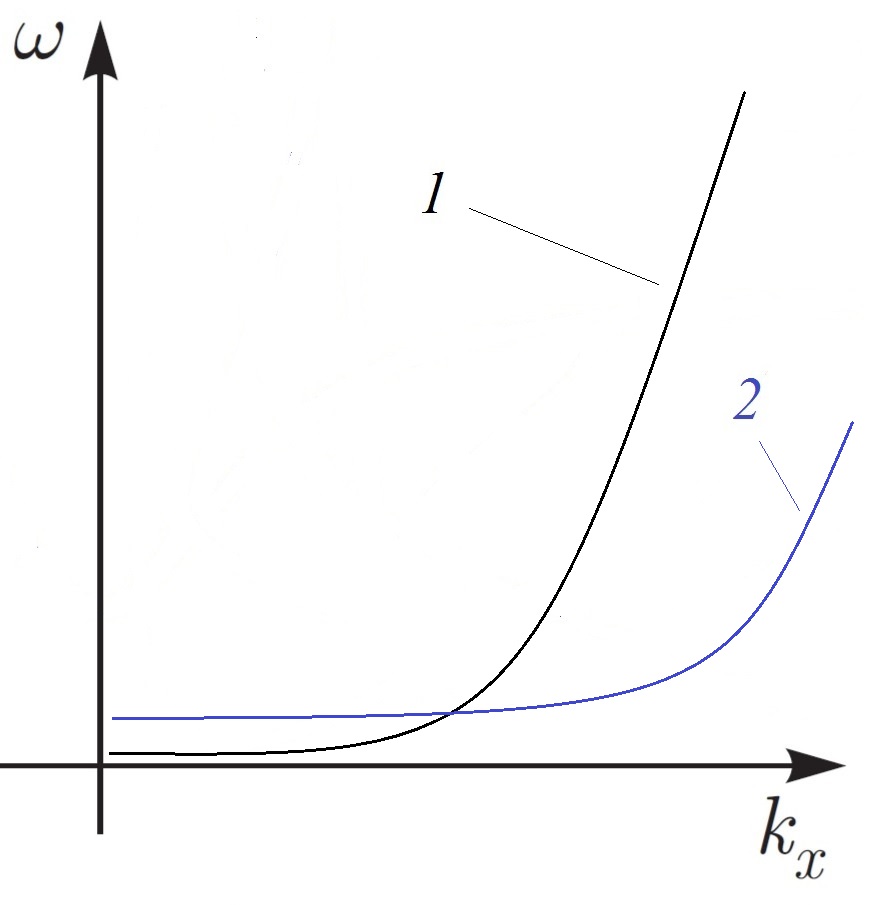}\]
\caption{The phase matching condition for magneto-Rossby waves in the absence of external magnetic field ($B_0=0$), 1: $\omega=\omega(k)$, 2: $\omega=\omega(k-k_{x_1})-\omega(k_{x_1})$}
\label{ris:experimcoded}
\end{minipage}
\end{center}
\end{figure}

To study three-wave interactions, we use the multiscale asymptotic method for the system of magnetohydrodynamic equations (\ref {18}) of stratified plasma in the approximation of two-layer shallow water on the beta plane in an external magnetic field  \cite{14}-\cite{16}. Since this method is widely used to study weakly nonlinear interactions, we restrict ourselves to a brief statement of the derivation of amplitude equations and give the expressions obtained for the interaction coefficients. We represent the solution of the system of equations (\ref {18}) in the form of an asymptotic series in the small parameter $ \varepsilon $:
\begin{equation}\label{39}
\mathbf{q}=\mathbf{q}_0+\varepsilon\mathbf{q}_1+\varepsilon^2\mathbf{q}_2+... .\end{equation}
In (\ref{39}) $\vec{q}_0$ is a stationary solution of a complete system (\ref {18}); $\vec{q}_1$ is a linear solution in the form of a plane wave with a known law of dispersion for shallow water equations in an external magnetic field (\ref{25}), (\ref{27}) or law of dispersion for shallow water equations in horizontal magnetic field (\ref{35}), (\ref{37}). The term $ \vec {q} _2 $ describes quadratic nonlinearity effect. In the equation for $ \vec {q} _2 $ obtained in the second order with respect to a small parameter $ \varepsilon $, the right-hand side contains the resonant terms leading to a linear growth of the solution with time and coordinates. Thus, condition $ \varepsilon ^ 2q_2 \ll \varepsilon q_1 $ is violated on large scales. Therefore, to eliminate the influence of resonant terms, we introduce a slowly varying amplitude depending on the slow time and large linear scales. 

Let us introduce a solution in the form of three magneto-Rossby waves satisfying the phase matching condition (\ref {38}):
\begin{eqnarray}\label{44}\mathbf{q}_1=\mathbf{q}_1(T_1,X_1,Y_1)\exp{[i(\omega T_0-k_xX_0-k_yY_0)]}=\phi\mathbf{a}(\mathbf{k}_1)\exp{(i\theta_1)}+\nonumber\\
+\psi\mathbf{a}(\mathbf{k_2})\exp{(i\theta_2)}+\chi\mathbf{a}(\mathbf{k_3})\exp{(i\theta_3)}+c.c.,\end{eqnarray}
where $\phi,\psi,\chi$ are amplitudes of interacting waves, $\theta_i=-\omega(\vec{k}_i)T_0+k_{xi}X_0+k_{yi}Y_0$ are wave phases, $\vec{a}$ is a complex wave vector.

“Fast” variables $ (T_0, X_0, Y_0) $ are related to “slow” ones $ (T_1, X_1, Y_1) $ by the following relations:
\begin{equation}\label{41}\frac{\partial }{\partial t}=\frac{\partial}{\partial T_0}+\varepsilon\frac{\partial}{\partial T_1}\textsf{;  }\frac{\partial }{\partial x}=\frac{\partial}{\partial X_0}+\varepsilon\frac{\partial}{\partial X_1}\textsf{;  }\frac{\partial }{\partial y}=\frac{\partial}{\partial Y_0}+\varepsilon\frac{\partial}{\partial Y_1}\end{equation}

Let us substitute the solution (\ref {39}) in the system (\ref {18}), of the magnetohydrodynamic equations of the stratified plasma in the shallow water approximation, taking into account (\ref {41}) and (\ref {44}). In the second order of smallness, we obtain the linear inhomogeneous equations with resonances in the right-hand side. To exclude resonant terms, we use the compatibility condition, which consists in the orthogonality of the right-hand side of the kernel of a linear operator on the left-hand side of the system of equations. Thus, multiplying the right-hand side by the eigenvector of the linear operator on the left-hand side, we write out successively the terms proportional to $ e^ {i \theta_1} $, $ e^ {i \theta_2} $ and $ e^ {i \theta_3} $ . As a result, we obtain a system for three amplitudes of interacting packets of magneto-Rossby waves in two-layer shallow water approximation:
\begin{equation}s_1\phi=f_1\psi^*\chi,\label{45.1}\end{equation}
\begin{equation}s_2\psi=f_2\phi^*\chi,\label{45.2}\end{equation}
\begin{equation}s_3\chi=f_3\phi\psi,\label{45.3}\end{equation}
where $ s_n $ is a differential operator with respect to the "slow" variables $ T_1, X_1, Y_1 $, and coefficients $ f_i $ depend only on the initial conditions and characteristics of interacting waves. The system (\ref {45.1}) - (\ref {45.3}) describes three-wave interactions of magneto-Rossby waves that satisfy the phase matching condition (\ref {38}). Each of the three equations of the system describes the nonlinear influence of the magnitudes of the amplitudes of the two interacting waves on the third wave.

In an external vertical magnetic field the differential operator $ s_n $ is following:
\[s_n=r_n\frac{\partial}{\partial T_1}+p_n\frac{\partial }{\partial X_1}+q_n\frac{\partial}{\partial Y_1},\]
where the coefficient $ r_n $ with the derivative for the slow time $ T_1 $ has the form:
\begin{eqnarray}\label{rn}
r_n=\left\{z_1a_1+h_{01}(ik_{y_n}z_2a_3+z_3a_4+z_4a_5+z_5a_6)\right\}+\nonumber\\
+\left\{z_6a_{2-1}+\Delta h_0(ik_{y_n}z_7a_7+z_8a_8+z_9a_9+z_{10}a_{10})\right\},\end{eqnarray}
the coefficient $ p_n $ with the coordinate derivative $ X_1 $ has the form:
\begin{equation}\label{pn}
p_n=\left\{h_{01}(z_1a_3+z_2ik_{y_n}ga_1)\right\}+\frac{\rho_2}{\rho_1}z_2ik_{y_n}gh_{01}a_{2-1}+\left\{\Delta h_0(z_6a_7+z_7ik_{y_n}ga_1)\right\},
\end{equation}
and the coefficient $ q_n $ with the coordinate derivative $ Y_1 $ is:
\small
\begin{eqnarray}\label{qn}q_n=\left\{h_{01}(z_1a_4+z_2(-i\omega(\mathbf{k}_n)a_3+ik_{x_n}ga_1-\frac{B_0a_5}{h_{01}}-f_0a_4)+z_3ga_1)\right\}+\nonumber\\
+\frac{\rho_2}{\rho_1}a_{2-1}gh_{01}(z_2ik_{x_n}+z_3)+\left\{\Delta h_0(z_6a_8+z_7(-i\omega(\mathbf{k}_n)a_7+ik_{x_n}ga_1-\frac{B_0a_9}{\Delta h_0}-f_0a_8)+z_8a_1)\right\}.
\end{eqnarray}
\normalsize
In these coefficients $\mathbf{a}=\mathbf{a}(\mathbf{k}_n)$, $n=1,2,3$. Here and after $a_{2-1}=a_2-a_1$. $n=1,2,3$.

The coefficients $ f_m $, depending only on the initial conditions and characteristics of the interacting waves, have the following form:
\begin{equation}\label{fm}
f_m=\left\{L_1\right\}+\frac{\rho_2}{\rho_1}R+\left\{L_2\right\}
\end{equation}
The expression $L_1$ corresponds to the bottom layer of plasma:
\[L_1=2iz_1[k_{x_m}a_1a_3+k_{y_m}a_1a_4]+z_2[k_{y_m}(2\omega(\mathbf{k}_m)a_1a_3-2k_{x_m}(a_3^*a_3-a_5^*a_5)-2k_{y_m}(a_3a_4-a_5a_6)+\]
\[+\alpha g(k_{x_l}a_{1}a^*_1-\alpha k_{x_c}a^*_{1}a_{1}))-if_0(k_{y_c}a^*_1a_{4}-k_{y_l}a_1a^*_{4})-2\beta a_1a_4]+\]
\[+z_3[-2i\omega(\mathbf{k}_m)a_1a_4+2ik_{y_m}(a_4^*a_4-a^*_6a_6)+2ik_{x_m}(a_3a_4-a_5a_6)+ik_{y_m}ga_{1}a_{1}^*+2f_0a_1a_3]-\]
\[-2iz_4[\omega(\mathbf{k}_m)a_1a_5+k_{y_m}h_0(a_3a_6-a_4a_5)]-2iz_5[\omega(\mathbf{k}_m)a_1a_6+k_{x_m}h_0(a_4a_5-a_3a_6)]\]
The expression $L_2$ correpsonds to the top layer of plasma:
\[L_2=2iz_6[k_{x_m}a_{2-1}a_7+k_{y_m}a_{2-1}a_8]
+z_7[k_{y_m}(2\omega(\mathbf{k}_m)a_{2-1}a_7-2k_{x_m}(a_7^*a_7-a_9^*a_9)-\]
\[-2k_{y_m}(a_7a_8-a_9a_{10})+\alpha g(k_{x_l}a_{2-1}a^*_{1}-\alpha k_{x_c}a^*_{2-1}a_{1}))-if_0(k_{y_c}a^*_{2-1}a_{8}-k_{y_l}a_{2-1}a^*_{8})-2\beta a_{2-1}a_8]+\]
\[+z_8[-2i\omega(\mathbf{k}_2)a_{2-1}a_8+2ik_{y_m}(a_8^*a_8-a_{10}^*a_{10})+2ik_{x_m}(a_7a_8-a_9a_{10})+ig(k_{y_c}a_{1}a^*_{2-1}-k_{y_l}a^*_{1}a_{2-1})+\]
\[+2f_0a_{2-1}a_7]-2iz_9[\omega(\mathbf{k}_m) a_{2-1}a_9+k_{y_m}\Delta h_0(a_7a_{10}-a_8a_9)]-2iz_{10}[\omega(\mathbf{k}_m)a_{2-1}a_{10}+\]
\[+k_{x_m}\Delta h_0(a_8a_9-a_7a_{10})]\]
The expression $R$ describes the influence of the stratification:
\[R=g[z_2\alpha k_{y_m}(k_{x_l}a_1a^*_{2-1}-\alpha k_{x_c}a^*_1a_{2-1})+z_3i(k_{y_c}a^*_{1}a_{2-1}-k_{y_l}a^*_{2-1}a_{1})]\]
Here compositions $a_ia_j=[a_i^*(\mathbf{k_l})a_j(\mathbf{k}_c)+a_i(\mathbf{k}_c)a_j^*(\mathbf{k}_l)]/2$. If index $m=1$, then index $l=2$, index $c=3$ and $\alpha=1$. If index $m=2$, then index $l=1$, index $c=3$ and $\alpha=1$. If index $m=3$, then index $l=1$, index $c=2$, $\alpha=-1$ and $a_i^*=a(\mathbf{k}_1)$.

Let us analyze the interaction coefficients (\ref {pn}) - (\ref {fm}) in more detail. The terms in each coefficient are divided into two similar in appearance expressions in curly brackets and middle terms, that include the ratio of plasma layers densities. The terms in the first curly bracket refer to the bottom layer of plasma, while the terms in the second curly bracket refer to the top one. Middle terms describe the effects of stratification. Assuming that one of the layer heights is zero, the densities are equal to each other ($ \rho_2 = \rho_1 $) and the component $ a_{2-1}$ of the complex wave vector $ \mathbf {a} $ is equal to  component $a_1$, coefficients (\ref {pn}) - (\ref {fm}) of interactions of three magnetic Rossby waves in an external magnetic field in the approximation of two-layer shallow water transform into coefficients for three interacting magnetic Rossby waves in an external vertical magnetic field in single-layer shallow water \cite{15}.

For the equations of a stratified plasma in two-layer shallow water approximation in the absence of an external magnetic field for the initial system, we obtain a similar set of three equations with a difference in the coefficients $ p_n $ and $ q_n $:
\[p_n'=\left\{h_{01}(z_1a_3+z_2ik_{y_n}(ga_1-a_5B_{x_{01}})-B_{x_{01}}(z_3a_6+z_4a_3+z_5a_4))\right\}+\frac{\rho_2}{\rho_1}z_2ik_{y_n}gh_{01}a_{2-1}+\]
\[+\left\{\Delta h_0(z_6a_7+ik_{y_n}z_7(g a_1-a_9B_{x_{02}})-B_{x_{02}}(z_8a_{10}+z_9a_7+z_{10}a_8))\right\}\]
\[q'_n=\left\{h_{01}(z_1a_4+z_2[-i\omega(\mathbf{k}_n) a_3-ia_5(B_{x_{01}}k_{x_n}+2B_{y_{01}}k_{y_n})+ik_{x_n}ga_1-f_0a_4]+\right.\]
\[\left.+z_3g(a_1-a_6B_{y_{01}})-z_4a_3B_{y_{01}}-z_5a_4B_{y_{01}})\right\}+\frac{\rho_2}{\rho_1}a_{2-1}h_{01}g(z_2ik_{x_n}+z_3)+\]
\[+\{\Delta h_0(z_6a_8+z_7[-i\omega a_7-ia_9(B_{x_{02}}k_{x_n}+2B_{y_{02}}k_{y_n})+ik_{x_n}ga_1-f_0a_8]+\]
\[+z_8(a_1-a_{10}B_{y_{02}})-z_9a_7B_{y_{02}}-z_{10}a_8B_{y_{02}}\}\]
For these expressions, as in the case of an external vertical magnetic field, the terms in the first curly bracket refer to the bottom layer, the terms in the second curly bracket refer to the top layer, and the middle terms decribe the effects of stratification. Since the remaining coefficients will differ from those similar to the external vertical magnetic field only by the components of the eigenvector $ \mathbf {z} $, then all previous conclusions are valid for them, including the transition to similar coefficients in single-layer shallow water \cite {15}. 

The system of equations (\ref {45.1}) - (\ref {45.3}) describes parametric instabilities in both cases considered, both in an external vertical magnetic field and in a horizontal field \cite{15}, \cite{16}. Since the equations of three-wave interactions for a stratified fluid in the two-layer approximation differ only in the interaction coefficients, the same parametric instabilities that were found in \cite {15} are realized in the two-layer model. The main difference in our case is in the increments of parametric instabilities and threshold values, which now depend on the ratio of densities.

In the case when the amplitude of one of the three waves at the initial moment is much larger than amplitudes of two other waves ($ \phi = \phi_0 \gg \psi, \chi $) the system (\ref {45.1}) - (\ref {45.3}) becomes:
\[s_2\psi=f_2\phi_0^*\chi\]
\[
s_3\chi=f_3\phi_0\psi\]
Finding the solution of the system in following form $(\psi,\chi)^T=(\psi',\chi)^T\exp(\Gamma_1T_1)$, we get the increment of this instability: $\Gamma_1=\sqrt{\frac{|f_2f_3|}{|r_2r_3|}}|\phi_0|>0$.

This is the case of the decay of a magneto-Rossby wave with a wave vector $ \vec {k}_1 $ and a frequency $ \omega (\vec {k}_1) $ into two magneto-Rossby waves with wave vectors $ \vec {k}_2 $, $ \vec {k}_3 $ and frequencies $ \omega (\vec {k}_2) $ and $ \omega (\vec {k}_3) $ respectively.

Let us consider the opposite situation - amplification the magneto-Rossby wave with wave vector $ \vec {k}_1 $ and frequency $ \omega (\vec {k}_1) $ by magneto-Rossby waves with wave vectors $ \vec {k}_2 $, $ \vec {k}_3 $ and frequencies $ \omega (\vec {k}_2) $, $ \omega ( \vec {k}_3) $ respectively. This type of instability occurs when the amplitude of one of the interacting waves is much smaller than amplitudes of the two other waves ($ \phi \ll \psi = \psi_0, \chi = \chi_0 $0. Then the system (\ref {45.1})-(\ref {45.3}) reduces to the equation:
\[s_1\phi=f_1\psi_0^*\chi_0.\]
Finding the solution of the equation $\phi=\phi'\exp{(\Gamma_2T_1)}$ we get the amplification coefficient of such an instability: $\Gamma_2=\frac{|f_1|}{|r_1|}|\psi_0\chi_0|>0$.

In the case of parametric decay with linear damping, we write the system (\ref {45.1}) - (\ref {45.3}) in the following form:
\[s_2\psi+\eta_2\psi=f_2\phi_0^*\chi\]
\[s_3\chi+\eta_3\chi=f_3\phi_0\psi\]
where $\eta_2\psi$, $\eta_3\chi$ are attenuation terms. In this case the threshold value of the amplitude of the pump wave is following: $\phi_{0_{cr}}=\sqrt{\frac{\eta_2\eta_3|r_{2}r_{3}|}{|f_{2}f_{3}|}}$. From this value the instability with increment $\Gamma_1'=\sqrt{\frac{|f_{2}f_{3}|}{|r_{2}r_{3}|}}\phi_{0_{cr}}$ starts. 

Similarly, for the case of parametric amplification, the system (\ref {45.1}) - (\ref {45.3}) reduces to the equation:
\[s_{v1}\phi+\eta_1\phi=f_{1}\psi_0^*\chi_0\]
where $ \eta_1 \phi $ is an attenuation term. In this case the threshold value of the product of amplitudes of the pump waves is following: $(\psi^*_0\chi_0)^{cr}=\eta_1\frac{|r_{1}|}{|f_{1}|}$. From this value the instability with increment $\Gamma_2'=\frac{|f_{1}|}{|r_{1}|}(\psi^*_0\chi_0)^{cr}$ starts.
\section{Conclusion.}
A nonlinear theory of flows of a thin layer of a stratified plasma in a field of gravity with a free-surface in an external vertical magnetic field in the presence of rotation is developed. Magnetohydrodynamic equations are obtained in the shallow water approximation when plasma is divided into two layers of different densities. In the particular case of equal heights and densities of each layer magnetohydrodynamic equations of two-layer shallow water are reduced to the equations of single-layer shallow water in an external magnetic field, obtained in \cite{9}, \cite{14}, \cite{36}. It is shown that despite the two-component and two-dimensional velocity fields in each of the layers, the magnetic field is three-component and axisymmetric in the shallow water approximation. The developed theory is generalized to the case of spherical flows in the beta-plane approximation for the Coriolis force. It is shown that in the linear approximation, the resulting system allows rotation for the magneto-Rossby waves, whose characteristics are modified by the ratio of the densities of plasma layers. It is found that frequency modifications associated with stratification reduce the group velocities of the magnetic Rossby waves in an external vertical magnetic field and increase their phase velocities.

The obtained system of magnetohydrodynamic equations of a stratified plasma in the two-layer shallow water approximation in the absence of an external magnetic field has a solution in the form of a stationary toroidal-poloidal magnetic field. The dispersion equation in this approximation also has a solution in the form of magneto-Rossby waves, modified by the density ratio. It was found that the frequency modification associated with stratification also reduces group velocities of magneto-Rossby waves. Phase velocity has a more complex behavior: for very small wave vectors ($ k_x <1 $) the frequency modification increases the phase velocity, while decreasing it for large values of the wave vector.

A qualitative analysis of the obtained dispersion relations shows the satisfaction of phase matching condition for three interacting waves, both in the presence of an external magnetic field and in its absence. Equations of three waves interactions are derived using the assymptotic multiscale method in both considered cases. The possibility of parametric instabilities is shown and their characteristics are found. Despite the universality of the obtained equations of three-wave interactions, the coefficients of interaction as well as characteristics of the found parametric instabilities are different for each considered case.

This work was supported by the Foundation for Development of Theoretical Physics and Mathematics “Basis” and the Grant RFBR (19-02-00016); completed under project KP19-270 “Issues of the origin and evolution of the Universe using ground-based observation and space research methods” of a program of large-scale projects for conducting fundamental research in priority areas defined by the Presidium of the Russian Academy of Sciences.
\newpage
\addcontentsline{toc}{section}{References}
\selectlanguage{english}

\end{document}